\documentclass[letterpaper]{article} 
\usepackage{aaai2026}  
\usepackage{times}  
\usepackage{helvet}  
\usepackage{courier}  
\usepackage[hyphens]{url}  
\usepackage{graphicx} 
\usepackage{natbib}  
\usepackage{caption} 
\usepackage{algorithm}
\usepackage{algorithmic}
\usepackage{url}

\usepackage{booktabs} 
\usepackage{multirow} 
\usepackage{subcaption}
\usepackage{makecell}
\usepackage{amsmath, amsfonts}

\usepackage{newfloat}
\usepackage{listings}
\DeclareCaptionStyle{ruled}{labelfont=normalfont,labelsep=colon,strut=off} 
\floatstyle{ruled}
\newfloat{listing}{tb}{lst}{}
\floatname{listing}{Listing}
\title{CycleChemist: A Dual-Pronged Machine Learning Framework for Organic Photovoltaic Discovery}
\author {
    Hou Hei Lam\equalcontrib \textsuperscript{\rm 1},
    Jiangjie Qiu\equalcontrib \textsuperscript{\rm 1}, 
    Xiuyuan Hu\textsuperscript{\rm 1}, 
    Wentao Li\textsuperscript{\rm 1}, 
    Fankun Zeng\textsuperscript{\rm 1}, 
    Siwei Fu\textsuperscript{\rm 1}, 
    Hao Zhang\textsuperscript{\rm 1}, 
    Xiaonan Wang\textsuperscript{\rm 1}
}
\affiliations {
    \textsuperscript{\rm 1}Tsinghua University\\
    \texttt{\{linhx22, qjj21, huxy22, liwt24, zfk23, fsw22\}@mails.tsinghua.edu.cn}\\
    \texttt{\{haozhang, wangxiaonan\}@tsinghua.edu.cn}
}

\usepackage{bibentry}

\begin{document}
\nocopyright
\maketitle

\begin{abstract}
Organic photovoltaic (OPV) materials offer a promising pathway for sustainable energy generation. However, their development is hindered by the challenge of identifying high-performance donor-acceptor pairs with optimal power conversion efficiencies (PCEs). Most existing design strategies focus exclusively on either the donor or the acceptor, rather than employing a unified model capable of designing both components. In this work, we introduce a dual-pronged machine learning framework for OPV discovery, integrating predictive modeling and generative molecular design. In this study, we propose the newly curated Organic Photovoltaic Donor-Acceptor Dataset (OPV\textsuperscript{2}D), the largest of its kind, comprising 2,000 experimentally characterized donor-acceptor pairs. This dataset serves as a comprehensive foundation for model training and evaluation. To enable accurate property prediction in organic photovoltaic (OPV) materials, we first introduce the Organic Photovoltaic Classifier (OPVC) to predict the likelihood that a given material exhibits OPV behavior. Complementing this, we develop a hierarchical graph neural network framework that integrates multi-task learning and cross-modal donor–acceptor interaction modeling. This framework includes the Molecular Orbital Energy Estimator (MOE\textsuperscript{2}) for predicting the highest occupied molecular orbital–lowest unoccupied molecular orbital (HOMO–LUMO) energy levels, and the Photovoltaic Performance Predictor (P\textsuperscript{3}) for estimating power conversion efficiency (PCE). In addition, we introduce the Material Generative Pretrained Transformer (MatGPT) to generate synthetically accessible organic semiconductors. Building on this, we propose a reinforcement learning strategy with three-objective policy optimization that guides molecular generation while preserving chemical validity. By bridging molecular representation learning with device performance prediction, our framework advances computational OPV material discovery. 
\end{abstract}

\begin{links}
     \link{Code}{https://github.com/HouHei0416/CycleChemist}
     \link{Datasets}{https://github.com/sunyrain/OPV2D}
\end{links}

\section{Introduction}

Artificial intelligence (AI) is increasingly recognized as a transformative force for advancing social good, especially in the field of renewable energy. Its integration with materials science is revolutionizing the discovery and design of functional materials through data-driven approaches that accelerate innovation. In the realm of sustainable energy, organic photovoltaic (OPV) materials have garnered significant interest due to their flexibility, lightweight nature, and low cost. However, despite these advantages, the development of high-performance OPV systems remains a major challenge, particularly in designing optimal donor–acceptor combinations. Traditional design methods typically rely on incremental modifications of existing molecular structures and extensive trial-and-error experimentation. This approach not only demands considerable human and material resources but also limits the exploration of the design space, making it difficult to identify potential high-performance molecules beyond current systems. Moreover, most molecular design strategies focus exclusively on either the donor or the acceptor, typically optimizing one while fully fixing a specific instance of the other. As a result, there is a lack of a generalizable model capable of designing both components in a flexible manner. 

\begin{figure*}[h]
    \centering
    \includegraphics[width=0.7\textwidth]{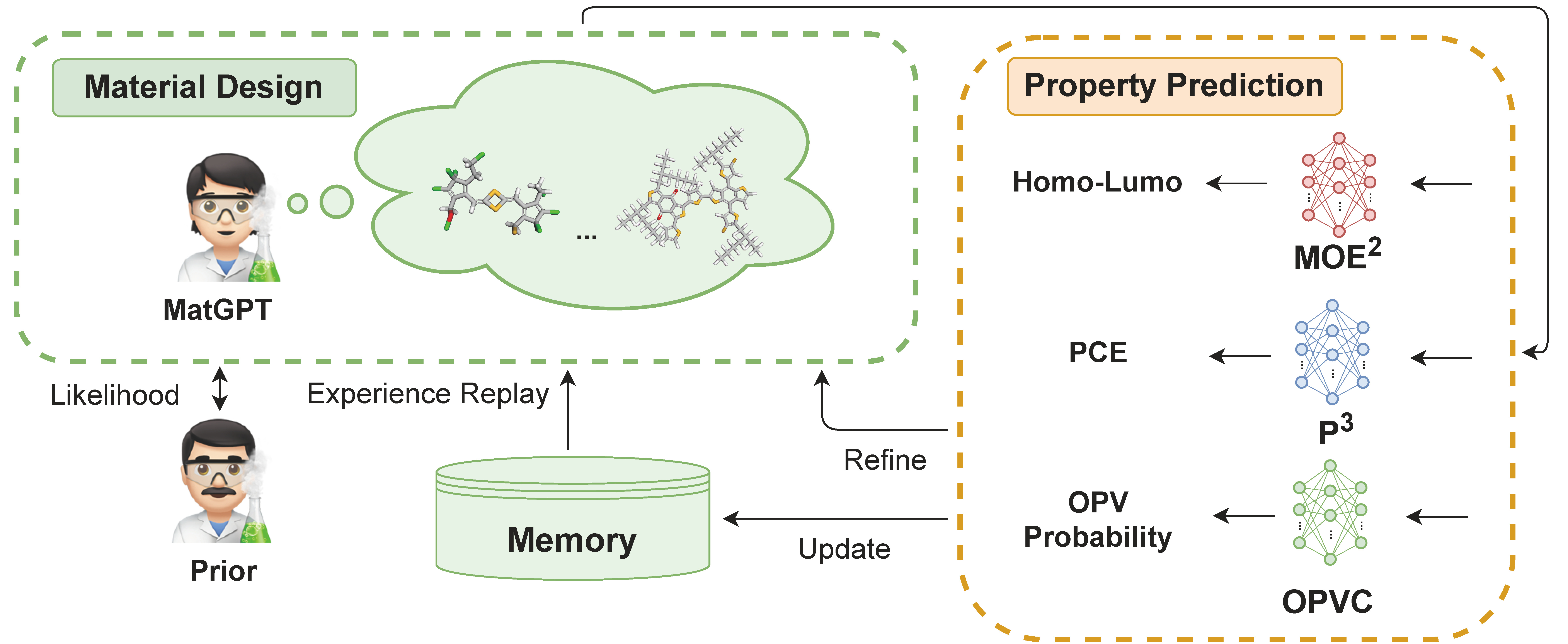}
    \caption{ 
The dual-pronged OPV materials discovery framework of our work integrates material generation, property prediction, and reinforcement learning for iterative enhancement. \textbf{Molecular Orbital Energy Estimator (MOE\textsuperscript{2})} is pre-trained via masked language modeling (MLM) and fine-tuned for HOMO-LUMO prediction. \textbf{Photovoltaic Performance Predictor (P\textsuperscript{3})} predicts power conversion efficiency (PCE) through supervised regression. \textbf{Organic Photovoltaic Classifier (OPVC)} is a Random Forest model that predicts the probability of a material being an organic photovoltaic (OPV). \textbf{Material Generative Pretrained Transformer (MatGPT)} is pre-trained using causal language modeling (CLM) and fine-tuned with reinforcement learning (RL) for high-PCE OPV molecular generation.}
    \label{fig:pipeline}
\end{figure*}

These limitations have motivated us to explore novel machine learning techniques for more efficient OPV material discovery. As shown in Figure~\ref{fig:pipeline}, we develop CycleChemist, a dual-pronged machine learning framework that integrates predictive modeling and generative molecular design, thereby overcoming previous paradigm limitations. The primary contributions are:

\begin{itemize}
    \item \textbf{Dataset Construction:} We curated the \textbf{Organic Photovoltaic Donor-Acceptor Dataset (OPV\textsuperscript{2}D)}, consisting of 2,000 experimentally characterized donor–acceptor pairs with calibrated properties, making it one of the largest OPV datasets available.
    
   \item \textbf{OPV Property Prediction Models:} We introduced \textbf{Organic Photovoltaic Classifier (OPVC)} that predicts the likelihood of a material exhibiting OPV behavior. In addition, our multi-task Graph Neural Network (GNN) predicts HOMO-LUMO levels and power conversion efficiency (PCE). It integrates the \textbf{Molecular Orbital Energy Estimator (MOE\textsuperscript{2})} for HOMO-LUMO predictions and the \textbf{Photovoltaic Performance Predictor (P\textsuperscript{3})} for PCE estimation.
    
    \item \textbf{Generative Model Development:} We designed and developed the \textbf{Material Generative Pretrained Transformer (MatGPT)} model. By incorporating Rotary Position Embedding (RoPE), a Gated Linear Unit (GLU) variant, and a diversified sampling method, this model is able to generate valid and diverse molecules. 
    
    \item \textbf{Reinforcement Learning Optimization:} A novel \textbf{three-objective RL strategy} balances theoretical performance, molecular validity, and electronic property distribution, guiding the generation of optimal OPV candidates.

\end{itemize}

\section{Related Work}

\subsection{Prediction of Power Conversion Efficiency (PCE)}
Current machine learning models used for predicting the power conversion efficiency (PCE) of organic photovoltaic (OPV) materials remain relatively simplistic. Limitations in both algorithm design and dataset sizes have resulted in insufficient predictive accuracy and generalizability. For example, a comprehensive survey of various embedding methods and machine learning models was conducted using 558 data points, with tree-based algorithms achieving the highest predictive correlation (approximately 0.61) \cite{Beyond2024review}. However, such a small dataset may not adequately represent mainstream OPV material systems, thereby significantly affecting generalizability. In another study, a larger dataset was compiled and a random forest model was trained, reporting an \( R^2 \) value of 0.71 \cite{Saeki2021}. Nonetheless, the reliance on experimental data as model input renders this approach impractical for large-scale screening studies. In \cite{1060}, training was performed on 1,060 donor–acceptor pairs using various graph neural network algorithms, with the Graph Attention Network (GAT) achieving the best result (reported \( r \) value of 0.74). Additionally, the model presented in \cite{300} was trained on 300 donor–acceptor pairs and involved detailed quantum mechanical calculations for each molecule during dataset preparation, introducing new quantum mechanical descriptors. On this small dataset, the gradient boosting (GB) model achieved the best performance with a reported \( r \) value of 0.92. However, the complexity of the quantum mechanical (QM) calculations limits the model's potential for large-scale screening. Similarly, another study utilizing only 100 data points incorporated numerous QM descriptors and reported an \( r \) value of 0.918 \cite{100}. In summary, these methods are either trained on relatively small datasets or require complex QM calculations or additional experimental parameters, all of which limit their applicability for large-scale screening and may lead to performance that is somewhat contingent on chance.

\subsection{Photovoltaic Material Discovery}
Previous research has predominantly focused on predicting the properties of photovoltaic materials, and in recent years, some scholars have begun to explore donor–acceptor design based on machine learning. According to the literature, methods for exploring high-efficiency OPV materials mainly include strategies such as disassembling and recombining existing molecular structures, as well as employing variational autoencoders (VAE) and genetic algorithms (GA) \cite{Hutchison2023, Sun2024, ZhangSA2025}. These approaches largely rely on optimizing known molecules, thereby limiting their capacity to generate and explore entirely novel molecules. Some studies do not involve molecular design at all but instead exhaustively enumerate possible donor–acceptor combinations, exploring potential high-efficiency pairs within a predefined design space \cite{100}. Moreover, most related studies employ datasets comprising only 100 to 1,500 samples \cite{Saeki2021, Min2020, Hutchison2023}, covering relatively narrow material systems, which consequently restricts the diversity of molecular fragments that the models can explore.

\section{Methodology}
\subsection{Dataset Preparation}

Based on previous works \cite{Saeki2021, Min2020}, we integrated and curated multiple organic photovoltaic datasets to construct the \textbf{Organic Photovoltaic Donor-Acceptor Dataset (OPV\textsuperscript{2}D)}, comprising approximately 2,000 donor–acceptor pairs—one of the largest OPV datasets to date. Molecules are represented using \textbf{SMILES} (\textbf{S}implified \textbf{M}olecular \textbf{I}nput \textbf{L}ine \textbf{E}ntry \textbf{S}ystem) \cite{weininger1988smiles}. Duplicate entries were removed, and for pairs with varying PCEs due to fabrication differences, the highest reported value was retained to align with our model's goal of predicting maximum achievable PCE. Data formats were standardized through manual inspection and corrections using the original literature. To ensure consistency, simplified alkyl chains in some datasets were restored to their authentic forms. 

\begin{figure}[h]
    \centering
    \includegraphics[width=1\linewidth]{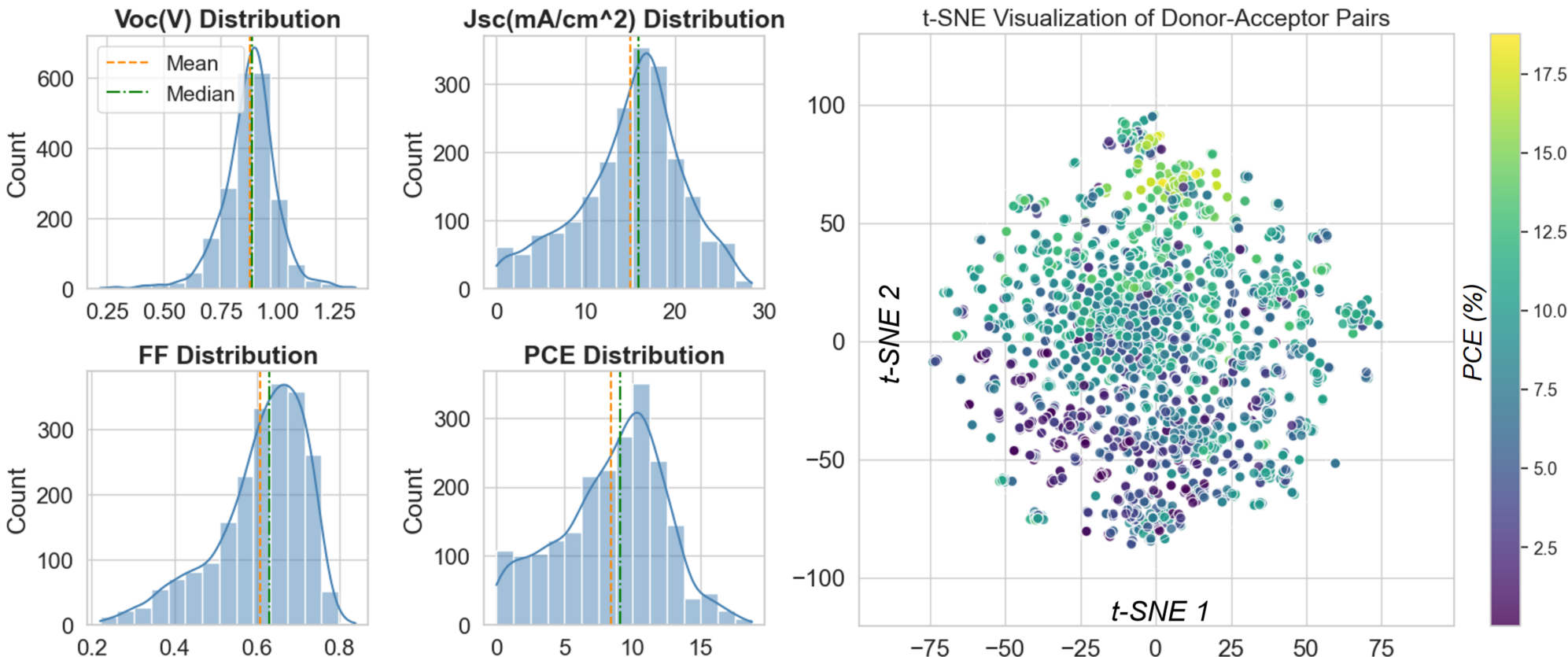}
    \caption{Dataset overview: photovoltaic parameter distributions and molecular structural diversity revealed by t-SNE visualization. The data spans a wide range of property values and structural variations.}
    \label{fig:visual}
\end{figure}

Figure~\ref{fig:visual} illustrates the statistical distribution and structural diversity of the photovoltaic parameters in our dataset. Histograms (left) for Voc (open-circuit voltage), Jsc (short-circuit current density), FF (fill factor), and PCE reveal central tendencies and spread. The t-SNE visualization (right) maps donor-acceptor pairs into a 2D latent space, color-coded by PCE. This underscores the broad coverage of the molecular configurations and experimental outcomes in our dataset, which supports robust model generalization.

\subsection{Property Prediction}
\begin{figure}[h]
    \centering
    \begin{subfigure}{\linewidth}
        \centering
        \includegraphics[width=\linewidth]{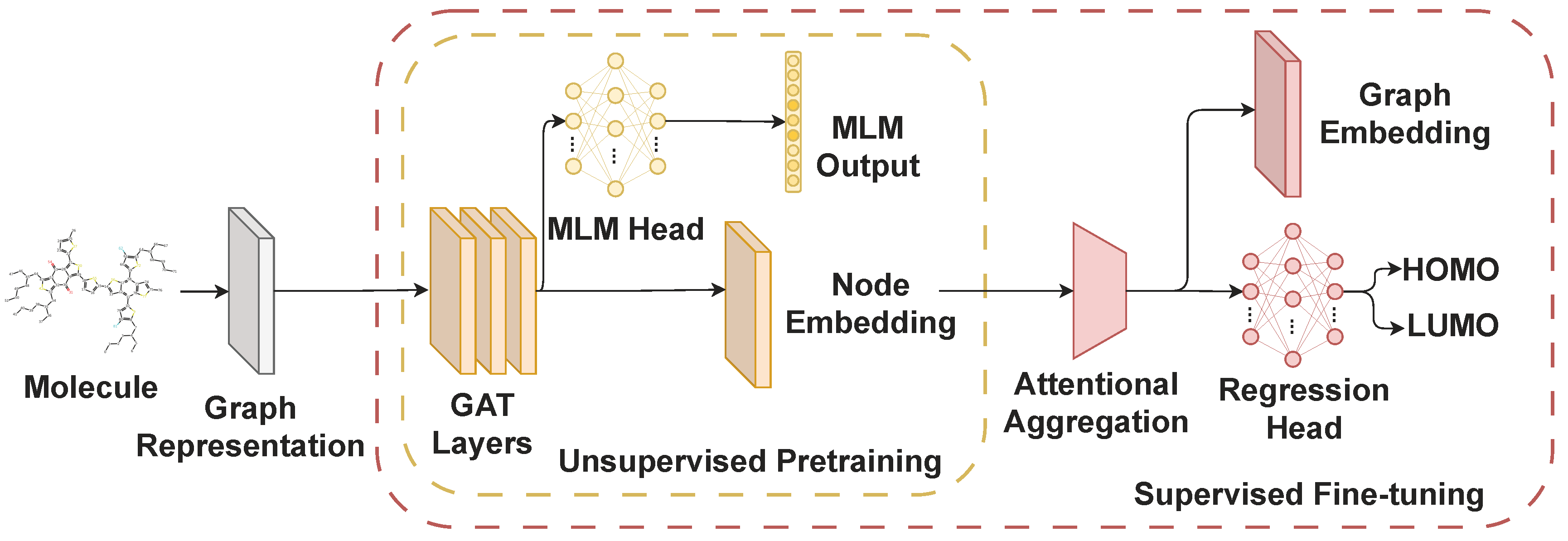}
        \caption{}
        \label{fig:moe2}
    \end{subfigure}
    \vskip\baselineskip
    \begin{subfigure}{\linewidth}
        \centering
        \includegraphics[width=\linewidth]{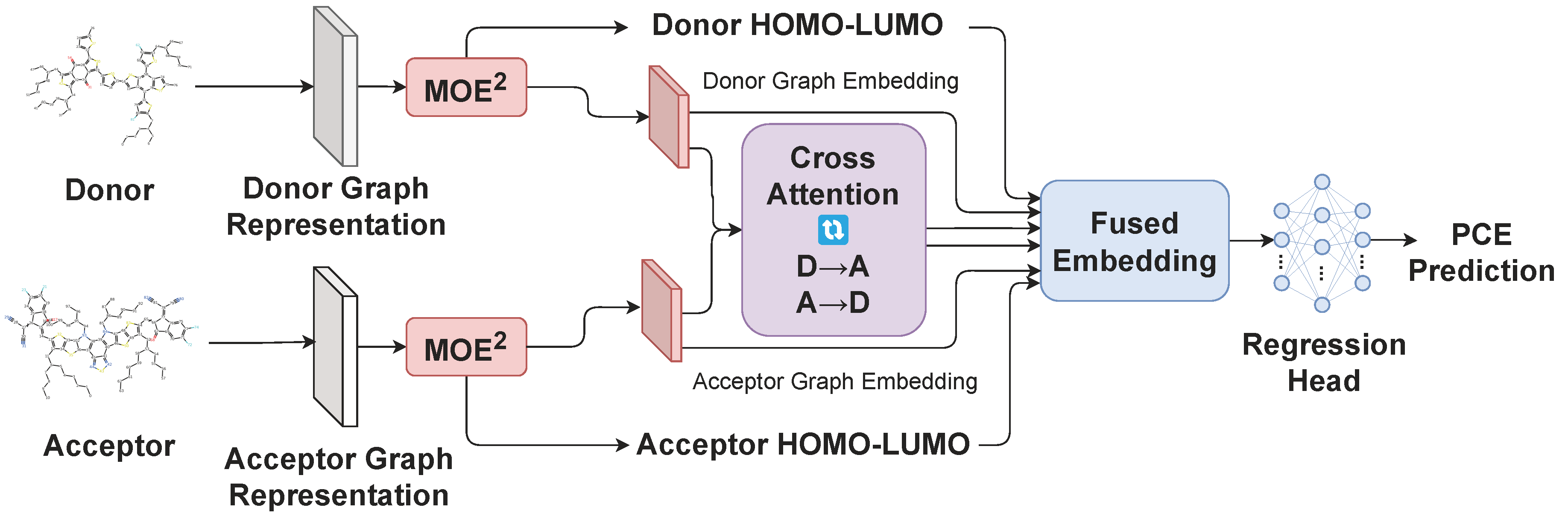}
        \caption{}
        \label{fig:p3}
    \end{subfigure}
    \caption{The architecture of (a) \textbf{MOE\textsuperscript{2}} and (b) \textbf{P\textsuperscript{3}} integrates hierarchical graph neural networks, multi-task learning, and cross-attention mechanisms to jointly predict molecular electronic properties (HOMO-LUMO) and photovoltaic performance (PCE), enabling efficient donor-acceptor screening for organic photovoltaics.}
    \label{fig:predictors}
\end{figure}

The hierarchical model integrates three complementary components to predict key aspects of molecular and device performance: (1) the Organic Photovoltaic Classifier (OPVC) for identifying OPV-active molecules, (2) the Molecular Orbital Energy Estimator (MOE\textsuperscript{2}) for predicting orbital energies HOMO-LUMO, and (3) the Photovoltaic Performance Predictor (P\textsuperscript{3}) for estimating power conversion efficiency (PCE) of donor–acceptor pairs (see Figure~\ref{fig:predictors}).

The design addresses three major challenges in OPV modeling: (1) representing complex molecular structures and electronic interactions, (2) learning transferable representations that capture both local and global chemical features, and (3) modeling donor–acceptor interactions governing charge transfer and exciton dynamics. These are tackled through a combination of cheminformatics features, molecular graph encoding, multi-task pretraining, and cross-attention mechanisms, balancing physical interpretability with predictive accuracy.

\subsubsection{OPV Classifier (OPVC)}

To estimate the likelihood of organic photovoltaic (OPV) activity, we train a binary Random Forest (RF) classifier \cite{breiman2001random}. The dataset comprises 1,500 OPV-active molecules from the OPV\textsuperscript{2}D dataset and 1,500 presumed-inactive molecules randomly selected from PubChem \cite{kim2025pubchem}. Each molecule is encoded using RDKit-derived molecular descriptors:

Given a molecular feature vector $\mathbf{x}$, the classifier outputs a probability score:
\begin{equation}
    P(\text{OPV} \mid \mathbf{x}) = \frac{1}{T_{tree}} \sum_{t_{tree}=1}^{T_{tree}} g_t(\mathbf{x})
\end{equation}
where $g_t(\mathbf{x}) \in \{0, 1\}$ is the prediction of the $t_{tree}$-th tree in the forest, and $T_{tree}$ is the total number of trees. This probability indicates the model's confidence that the molecule is OPV-active.

\subsubsection{Molecular Orbital Energy Estimator (MOE\textsuperscript{2})}
\paragraph{Graph Encoder Architecture}
The molecular encoder uses a three-stage GATv2-based hierarchy \cite{brody2021attentive} to capture local and global interactions. Feature propagation at each layer is defined as:
\begin{equation}
    \mathbf{z}_i^{(l+1)} = \text{GATv2}\left(\mathbf{z}_i^{(l)}, \bigoplus_{j \in \mathcal{M}(i)} \psi(\mathbf{z}_j^{(l)}, \mathbf{e}_{ij})\right),
\end{equation}
where, $\mathbf{z}_i^{(l)}$ denotes the feature vector of node $i$ at layer $l$, and $\mathcal{M}(i)$ represents its neighboring nodes. The operator $\bigoplus$ is an attention-based message aggregation function with 8 attention heads, and $\psi(\mathbf{z}_j^{(l)}, \mathbf{e}_{ij})$ incorporates edge attributes $\mathbf{e}_{ij}$ from molecular graph, which encode bond properties such as type, aromaticity, and conjugation. Molecular graphs are derived from SMILES using RDKit \cite{rdkit}, with atoms as nodes and bonds as edges. An edge-conditioned dynamic attention mechanism allows the model to distinguish between different interatomic interactions, such as covalent bonds, and $\pi$–$\pi$.

\paragraph{Attentional Graph Regression}
After that, the final prediction is implemented through a dedicated regression head operating on graph-level embeddings. For a molecular graph \( G \), the orbital energy prediction is computed as:
\begin{align}
    \mathbf{h}_G &= \text{AttentionalAggregation}(\mathbf{Z}),  \\
    \hat{y}_{HOMO}, \hat{y}_{LUMO} &= \mathbf{W}_2 \cdot \text{ReLU}(\mathbf{W}_1\mathbf{h}_G + \mathbf{b}_1) + \mathbf{b}_2, 
\end{align}
where \( \mathbf{Z} \in \mathbb{R}^{N_G \times d_G} \) represents node embeddings from the final GATv2 layer, where 
$N_G$ is the number of nodes in the molecular graph and $d_G$ is the embedding dimension of each node. The attentional aggregation generates graph embeddings \( \mathbf{h}_G \in \mathbb{R}^{d_G} \) through learnable attention weights over nodes, capturing key information relevant to frontier orbital prediction. The regression head consists of two linear layers with a hidden size of 128, ReLU activation, and dropout, followed by a final layer projecting to 2 outputs for simultaneous HOMO and LUMO energy prediction.

\subsubsection{Photovoltaic Performance Predictor (P\textsuperscript{3})}
\paragraph{Cross-Domain Interaction Module}
For a donor-acceptor pair \( (D, A) \), the interaction energy is modeled through a dual cross-attention mechanism:
\begin{align}
    \text{Attn}_D &= \text{softmax} \left(\frac{Q_A K_D^T}{\sqrt{d_k}}\right) V_D, \\
    \text{Attn}_A &= \text{softmax} \left(\frac{Q_D K_A^T}{\sqrt{d_k}}\right) V_A
\end{align}

Here, $Q_D, K_D, V_D$ and $Q_A, K_A, V_A$ are the query, key, and value projections of the donor and acceptor embeddings, $\mathbf{h}_D$ and $\mathbf{h}_A$, respectively. The key dimension $d_k$ is used to scale attention scores for numerical stability. This dual cross-attention mechanism captures both donor-to-acceptor charge transfer and acceptor-induced polarization by computing directional interaction representations for each.

\paragraph{Nonlinear Regression}
Final power conversion efficiency (PCE) prediction is computed via a non-linear fully connected feedforward regression function $\Psi$, integrating donor and acceptor features, cross-attention outputs, and predicted orbital energies through feature concatenation:
\begin{equation}
\begin{split}
    \hat{y}_{PCE} = \Psi(&[\mathbf{h}_D;\, \mathbf{h}_A;\, \text{Attn}_D;\, \text{Attn}_A; \\
                         &\hat{y}_{HOMO}^D;\, \hat{y}_{LUMO}^D;\, \hat{y}_{HOMO}^A;\, \hat{y}_{LUMO}^A]).
\end{split}
\end{equation}

The architecture maintains physical interpretability by: (1) separating molecular and interaction features, (2) using dedicated heads for energy prediction, and (3) enabling attention weights to reflect orbital overlap. This unified design supports simultaneous molecular and device-level prediction.
  
\subsection{MatGPT Architecture}  
The \textbf{Material Generative Pretrained Transformer (MatGPT)} extends the GPT-2 architecture~\cite{radford2019language}—a stack of masked transformer decoders—with targeted changes to improve molecular generation. Designed to prioritize both chemical validity and diversity, MatGPT learns from SMILES sequences to generate novel molecules. Our baseline implementation retains GPT's core components: 8 transformer layers, each containing 8 multi-head self-attention heads, followed by position-wise feed-forward networks. The model processes sequences of 340 tokens using 256-dimensional embeddings.

Two key architectural modifications improve the model’s ability to capture molecular structural constraints and enhance generation diversity. (1) We adopt Rotary Positional Encoding (RoPE) \cite{su2024roformer}, which replaces absolute positional embeddings with rotational transformations, enabling implicit encoding of cyclic and relative patterns common in molecular graphs. (2) We introduce a Gated Linear Unit (GLU)-based feed-forward layer that better captures local and global feature interactions. Specifically, the hidden representation $x \in \mathbb{R}^d$ is transformed as follows:
\begin{equation}
    \text{GLU}(x) = \sigma(W_{gate} x + b_{gate}) \odot \sigma(W_{value} x + b_{value}),
\end{equation}
where $\odot$ denotes element-wise multiplication and $\sigma$ is the GELU \cite{hendrycks2016gaussian} activation function. This formulation allows dynamic gating of features while maintaining computational efficiency.
To further improve molecular generation, we implement a diversified sampling method that combines top-$p$ filtering \cite{holtzman2019curious}, and temperature annealing. At each generation step, the logits are adjusted using a dynamic temperature schedule $T(t)$, interpolating between $T_{\text{max}}$ and $T_{\text{min}}$ over time:
\begin{equation}
    T(t) = T_{\text{min}} + (T_{\text{max}} - T_{\text{min}}) \cdot \left(1 - \frac{t}{T_{\text{total}}}\right).
\end{equation}
This approach encourages exploration during early steps and gradually sharpens token distributions as decoding proceeds. The combined sampling strategy allows for the generation of structurally diverse and chemically valid molecules, addressing the mode collapse problem often observed in autoregressive molecular models. 

\subsection{Reinforcement Learning Framework for Photovoltaic Material Discovery}

Our reinforcement learning framework employs a three-dimensional optimization strategy that combines direct photovoltaic performance maximization with statistical property alignment. The system utilizes a single MatGPT-based generative agent initialized from a pre-trained prior, guided by a composite loss function balancing theoretical performance, molecular validity, and electronic property distribution control.

The core optimization objective integrates three critical components into the total loss function, operating over generated molecules $u \sim P_{\text{Agent}}$:
\begin{equation}
\begin{aligned}
    \mathcal{L}_{\text{total}} = 
    &\underbrace{\mathbb{E}_u\left[\left(\zeta_1 \cdot r(u) - (\log P_{\text{Prior}}(u) - \log P_{\text{Agent}}(u))\right)^2\right]}_{\substack{\text{Photovoltaic Performance Optimization} \\ \text{and Prior-Policy Alignment}}} \\
    &+ \underbrace{\zeta_2 \cdot D_{\text{KL}}(\mathcal{N}_{\text{gen}} \parallel \mathcal{N}_{\text{data}})}_{\text{Property Distribution Control}},
\end{aligned}
\label{eq:loss}
\end{equation}
where $\zeta_1$ controls the reward scaling, determining how strongly the model is incentivized by the reinforcement learning signal. The term $(\log P_{\text{Prior}}(u) - \log P_{\text{Agent}}(u))$ enforces alignment between the agent’s learned policy and the prior distribution. The coefficient $\zeta_2$ regulates the property alignment strength, where positive values encourage distribution matching, while negative values promote divergence.

The reward function focuses on photovoltaic performance while maintaining chemical validity:
\begin{align}
r(u) =\ & \max\Big(0,\ 
    \min\big( 
        \hat{y}_{\text{PCE}}(u) \cdot P(\text{OPV} \mid \mathbf{x(u)}),\ 33.7 
    \big) 
\Big) \notag \\
& \cdot\ \mathbb{I}_{\text{valid}}(u),
\end{align}
where $\hat{y}_{\text{PCE}}$ is the predicted power conversion efficiency by P\textsuperscript{3}, $P(\text{OPV} \mid \mathbf{x(u)})$ is the OPV likelihood calculated by OPVC, and $\mathbb{I}_{\text{valid}}$ is the validity indicator function. It is constrained by the Shockley-Queisser theoretical efficiency limit~\cite{shockley2018detailed}. The KL divergence term provides statistical control over electronic properties:
\begin{equation}
D_{\text{KL}} = \sum_{\substack{
i \in \{\text{HOMO},\\ \text{LUMO}\}
}} 
\frac{1}{2} \left[ 
\log\frac{\sigma_i^2}{\tilde{\sigma}_i^2} 
+ \frac{\tilde{\sigma}_i^2 + (\tilde{\mu}_i - \mu_i)^2}{\sigma_i^2} - 1 
\right],
\end{equation}
where $(\mu_{\text{HOMO}}, \sigma_{\text{HOMO}})$ and $(\mu_{\text{LUMO}}, \sigma_{\text{LUMO}})$ represent reference distributions from OPV\textsuperscript{2}D experimental data, while $(\tilde{\mu}, \tilde{\sigma})$ denote generated molecule statistics, which are predicted by MOE\textsuperscript{2}.

This framework incorporates experience replay \cite{lin1992self}, strategically reintegrating the top-performing 5\% of historical molecules into subsequent training batches to reinforce successful chemical motifs. To maintain molecular diversity, a similarity threshold based on the Tanimoto index ($>$ 0.7) is applied to suppress redundant structures. 

\section{Experiments}
\subsection{Property Prediction}

\subsubsection{OPV Likelihood Prediction}
We trained OPVC on a dataset consists of 1,500 positive examples from the OPV\textsuperscript{2}D dataset and 1,500 negative samples randomly drawn from the PubChem database \cite{kim2025pubchem}, using 8:2 train-test split. The model was configured with 200 estimators, a maximum depth of 15, and class-balanced weighting. It achieved excellent performance, with a test accuracy of 99.16\% and F1 score of 0.9916. The AUC reached 0.9985 on the test set, indicating strong generalization and robust classification performance.

\subsubsection{HOMO-LUMO Prediction}
We pre-trained the graph encoder of MOE\textsuperscript{2} using three tasks: molecular mask reconstruction, computed HOMO-LUMO prediction, and literature-reported HOMO-LUMO prediction. For the first two tasks, we used over 51,000 non-fullerene acceptors (NFAs) from the reported dataset~\cite{Lopez201751k}, and for the third task, we used the proposed OPV\textsuperscript{2}D dataset. A 9:1 train-test split was employed for all tasks. The model achieved 99.997\% accuracy on the reconstruction task and $R^2$ scores of 0.9866 and 0.9784 for the computed and literature-reported HOMO-LUMO prediction tasks, respectively.

To ensure a fair comparison in the baseline experiments, we also pre-trained other graph encoders using the same three tasks and datasets. Three architectures—GCNConv~\cite{kipf2016semi}, GATv2Conv~\cite{brody2021attentive}, and GINConv~\cite{xu2018powerful}—were trained under identical hyperparameter settings. 

\subsubsection{PCE Prediction}
After pretraining, we evaluated the learned graph embeddings against traditional Morgan fingerprints on the OPV\textsuperscript{2}D dataset. For each sample, the embeddings were concatenated and used as input to a range of regression models, including Random Forest (RF) \cite{breiman2001random}, Support Vector Machine (SVM) \cite{cortes1995support}, Gradient Boosting (GB) \cite{friedman2001greedy}, XGBoost \cite{chen2016xgboost}, CatBoost \cite{prokhorenkova2018catboost}, LightGBM (LGBM) \cite{ke2017lightgbm}, Gaussian Processes (GP) \cite{seeger2004gaussian}, Histogram-based Gradient Boosting (HGB) \cite{pedregosa2011scikit}, and Natural Gradient Boosting (NGB) \cite{duan2020ngboost}, to predict power conversion efficiency (PCE). We used 5-fold cross-validation and the coefficient of determination ($R^2$) as the evaluation metric. Baseline results indicated that GAT-based embeddings, which incorporate edge weight information, consistently outperformed traditional approaches.

Building upon these GAT embeddings, we introduced P\textsuperscript{3}, a model that integrates self-attention and cross-attention mechanisms to more effectively capture intra- and inter-molecular interactions. This enhancement led to a significant improvement in performance, achieving a peak $R^2$ score of 0.716. To ensure fairness and reproducibility, we applied consistent hyperparameter configurations across all models. Performance results are summarized in Table~\ref{tab:r2_scores_horizontal}.

\begin{table}[ht]
\begin{center}
\setlength{\tabcolsep}{1mm}
\fontsize{9}{10.5}\selectfont
\caption{Average $R^2$ Scores for Models Predicting PCE using Morgan FP and Graph Neural Embeddings. Best values in \textbf{bold}, worst in {\textit{italic}} (horizontal comparison within each row). Error values are formatted in \raisebox{-0.1ex}{\scriptsize{±error}}.}
\label{tab:r2_scores_horizontal}

\begin{tabular}{l c c c c} 
\toprule
Model & Morgan FP & \multicolumn{3}{c}{Graph Embeddings} \\
\cmidrule(lr){3-5}
 &  & \makecell{GNN\\embedding} & \makecell{GAT\\embedding} & \makecell{GIN\\embedding} \\
\midrule
RF          & \textbf{0.691\raisebox{-0.1ex}{\scriptsize{±0.030}}} & 0.673\raisebox{-0.1ex}{\scriptsize{±0.038}} & 0.688\raisebox{-0.1ex}{\scriptsize{±0.023}} & {\textit{0.670\raisebox{-0.1ex}{\scriptsize{±0.031}}}} \\
SVM         & 0.577\raisebox{-0.1ex}{\scriptsize{±0.039}} & 0.585\raisebox{-0.1ex}{\scriptsize{±0.040}} & \textbf{0.611\raisebox{-0.1ex}{\scriptsize{±0.044}}} & {\textit{0.575\raisebox{-0.1ex}{\scriptsize{±0.046}}}} \\
GB          & 0.658\raisebox{-0.1ex}{\scriptsize{±0.036}} & 0.663\raisebox{-0.1ex}{\scriptsize{±0.038}} & \textbf{0.667\raisebox{-0.1ex}{\scriptsize{±0.022}}} & {\textit{0.640\raisebox{-0.1ex}{\scriptsize{±0.030}}}} \\
XGB         & {\textit{0.675\raisebox{-0.1ex}{\scriptsize{±0.036}}}} & 0.689\raisebox{-0.1ex}{\scriptsize{±0.041}} & \textbf{0.700\raisebox{-0.1ex}{\scriptsize{±0.027}}} & 0.682\raisebox{-0.1ex}{\scriptsize{±0.030}} \\
CatBoost    & 0.648\raisebox{-0.1ex}{\scriptsize{±0.031}} & 0.648\raisebox{-0.1ex}{\scriptsize{±0.037}} & \textbf{0.655\raisebox{-0.1ex}{\scriptsize{±0.037}}} & {\textit{0.633\raisebox{-0.1ex}{\scriptsize{±0.038}}}} \\
LGBM        & 0.689\raisebox{-0.1ex}{\scriptsize{±0.038}} & 0.687\raisebox{-0.1ex}{\scriptsize{±0.037}} & \textbf{0.703\raisebox{-0.1ex}{\scriptsize{±0.017}}} & {\textit{0.682\raisebox{-0.1ex}{\scriptsize{±0.033}}}} \\
GP          & 0.131\raisebox{-0.1ex}{\scriptsize{±0.085}} & {\textit{-0.366\raisebox{-0.1ex}{\scriptsize{±0.240}}}} & \textbf{0.301\raisebox{-0.1ex}{\scriptsize{±0.155}}} & -0.110\raisebox{-0.1ex}{\scriptsize{±0.168}} \\
HGB         & 0.686\raisebox{-0.1ex}{\scriptsize{±0.038}} & 0.686\raisebox{-0.1ex}{\scriptsize{±0.038}} & \textbf{0.704\raisebox{-0.1ex}{\scriptsize{±0.022}}} & {\textit{0.681\raisebox{-0.1ex}{\scriptsize{±0.032}}}} \\
NGB         & \textbf{0.568\raisebox{-0.1ex}{\scriptsize{±0.033}}} & 0.565\raisebox{-0.1ex}{\scriptsize{±0.040}} & 0.565\raisebox{-0.1ex}{\scriptsize{±0.034}} & {\textit{0.535\raisebox{-0.1ex}{\scriptsize{±0.041}}}} \\
\midrule
\textbf{P\textsuperscript{3} (Ours)}        & -- & -- & \textbf{0.736\raisebox{-0.1ex}{\scriptsize{±0.033}}} & -- \\
\bottomrule 
\end{tabular}

\end{center}
\end{table}

\subsection{Molecular Generator Performance}
To evaluate the molecular generation capability of MatGPT, we adopted the Molecular Sets (MOSES) \cite{polykovskiy2020molecular} benchmarking platform with protocol adaptations for organic photovoltaic (OPV) applications. The MOSES metrics were selected based on their ability to address three critical requirements for OPV material discovery: (1) \textit{Validity} to ensure synthesizability of donor/acceptor candidates, (2) \textit{Novelty} to avoid rediscovery of known low-efficiency materials, and (3) \textit{Diversity (IntDiv, IntDiv2)} to explore broad chemical spaces for synergistic pairings. Besides vanilla GPT-2, we compared MatGPT against baseline models spanning diverse generative paradigms, including VAE (autoencoder-based) \cite{gomez2018automatic}, ORGAN (Generative Adversarial Networks) \cite{guimaraes2017objective}, a combinatorial generator (template-based) \cite{polykovskiy2020molecular}, and NGram (rule-based) \cite{brown1992class}. 

\subsubsection{Experimental Setup}
We selected PubChem \cite{kim2025pubchem} for pretraining due to its broader chemical diversity compared to drug-focused datasets like ZINC Clean Leads \cite{sterling2015zinc} and ChEMBL \cite{gaulton2012chembl}. Domain-specific preprocessing was applied to tailor the data for OPV materials, filtering SMILES strings to 15–300 characters and excluding ionic or multi-molecule entries. From the cleaned data, 5 million molecules were sampled to train MatGPT-Large, while MatGPT, GPT-2, and other baselines used a 1 million subset. An 8:2 split was used for training and reference sets. MatGPT and GPT-2 were trained for 30 epochs on the Causal Language Modeling task \cite{radford2018improving} using a cosine annealing schedule (initial learning rate: 1e-4); other baselines followed MOSES defaults for fair comparison. Each model generated 30,000 molecules, evaluated with MOSES metrics. MatGPT used top-p sampling (p = 0.91) and temperature scaling (0.5–1.4) for diversity. All experiments ran on a single NVIDIA RTX 4090 GPU.
\begin{figure}[t]
    \centering
    \includegraphics[width=1\linewidth]{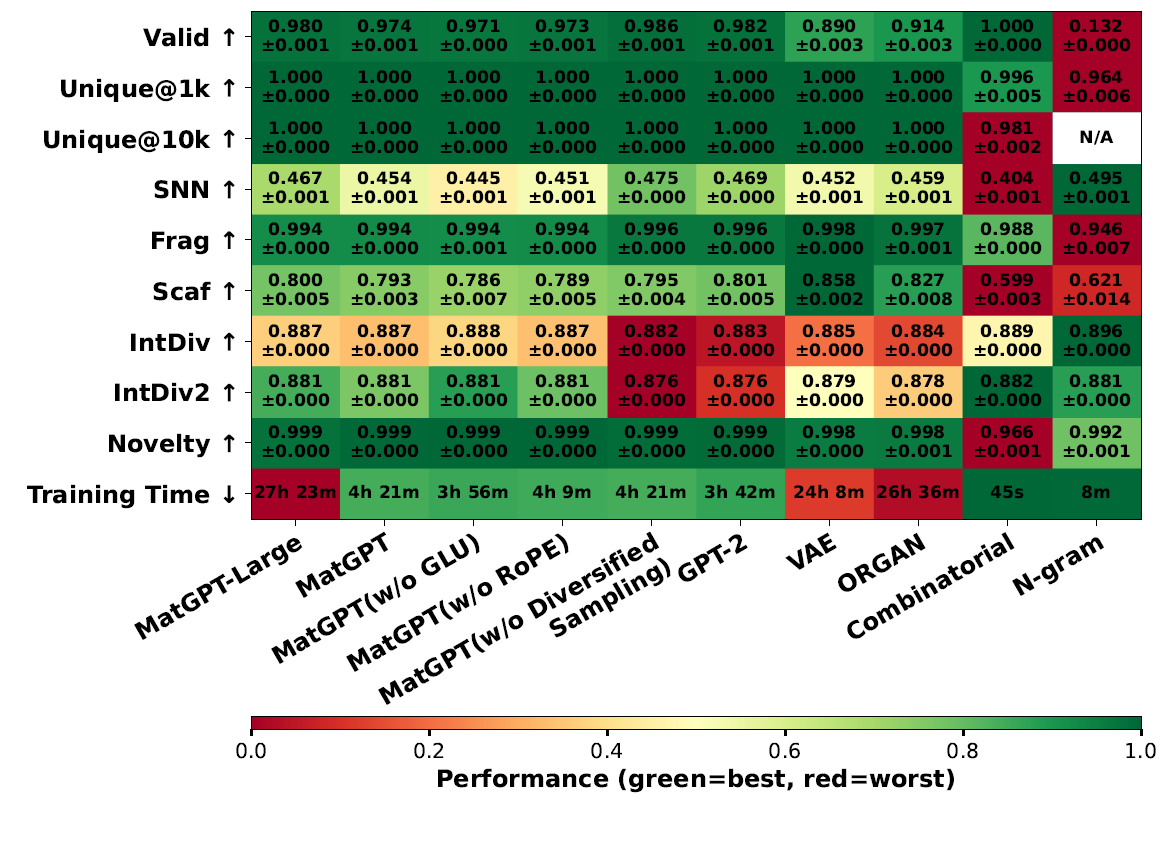}
    \caption{MOSES benchmark performance comparison. Arrows indicate optimal direction ($\uparrow$=higher better, $\downarrow$=lower better). Best values in green, worst in red. Values show mean ± std. dev. All models were trained on the 1M PubChem dataset, except for MatGPT-Large, which was trained on 5M. N-gram generated only around 3,200 valid molecules (insufficient for Unique@10k).}
    \label{fig:moses}
\end{figure}

\subsubsection{Experimental Results}
As shown in Figure~\ref{fig:moses}, MatGPT achieves a strong balance between validity (97.4\%) and novelty (99.9\%) while generating 100\% unique molecules, outperforming traditional baselines such as VAE (89.0\% validity) and ORGAN (91.4\%). Training on a larger dataset, MatGPT-Large further improves validity to 98.0\% without compromising diversity.

MatGPT maintains acceptable similarity to the nearest neighbor (SNN = 0.793), indicating that generated molecules stay close to the reference distribution. Its high fragment similarity (Frag = 0.994) suggests balanced representation of chemical fragments, while the scaffold similarity (Scaff = 0.793) reflects good alignment in core structural motifs. Additionally, MatGPT shows relatively high internal diversity (IntDivP) compared to other baselines, demonstrating its ability to generate a chemically diverse set of molecules.

Ablation studies highlight the importance of architectural components: removing RoPE or GLU layers slightly reduces validity, SNN, and Scaff, while disabling diversified sampling increases validity (98.6\%) but lowers diversity (IntDiv2 = 0.876). 

\subsection{Designing Donor and Acceptor Molecules}
We begin with the pre-trained MatGPT-Large model and fine-tune it on the OPV\textsuperscript{2}D dataset, updating only the gated linear unit (GLU) layers while freezing all other parameters. This is followed by RL–based fine-tuning, where all model parameters remain frozen except for the GLU layers. During RL, we use the trained OPVC to estimate OPV activity, MOE\textsuperscript{2} for predicting HOMO and LUMO energy levels, and P\textsuperscript{3} for predicting power conversion efficiency (PCE). The loss function includes a reward scaling coefficient $\zeta_1 = 100$. The RL fine-tuning was run for 500 steps with a batch size of 128 and a learning rate of $1 \times 10^{-4}$, completing in approximately 3 hours on a single NVIDIA RTX 4090 GPU.

\begin{figure}[t]
    \centering
    \includegraphics[width=1\linewidth]{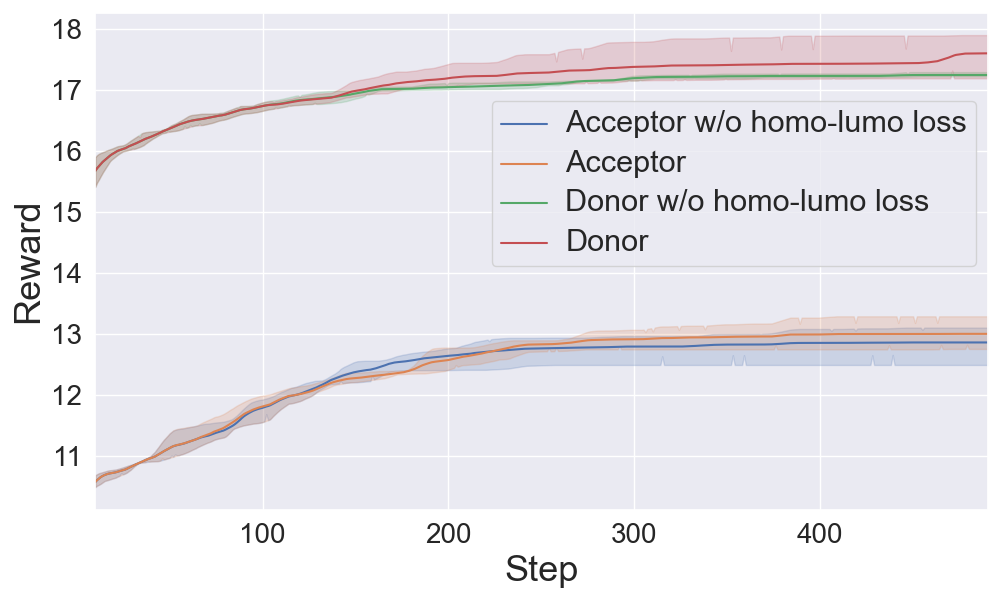}
    \caption{Mean top-10 reward in the memory during RL training.}\label{fig:training_curve}
\end{figure}

Figure~\ref{fig:training_curve} shows the mean top-10 reward in the memory over training steps for various reinforcement learning (RL) configurations. Two search settings are evaluated: fixing PTB7-Th (C\textsubscript{49}H\textsubscript{57}FO\textsubscript{2}S\textsubscript{6}) as the donor while searching for acceptors, and fixing Y6 (C\textsubscript{82}H\textsubscript{86}F\textsubscript{4}N\textsubscript{8}O\textsubscript{2}S\textsubscript{5}) as the acceptor while searching for donors. Each setting is tested without and with the Property Distribution Control term, controlled by the coefficient $\zeta_2 = 500$, to guide optimization toward OPV-favorable molecules based on OPV\textsuperscript{2}D. With the Property Distribution Control term enabled, both settings consistently achieve higher mean rewards, indicating improved generation of OPV-favorable candidates.

\begin{figure}
    \centering
    \includegraphics[width=1\linewidth]{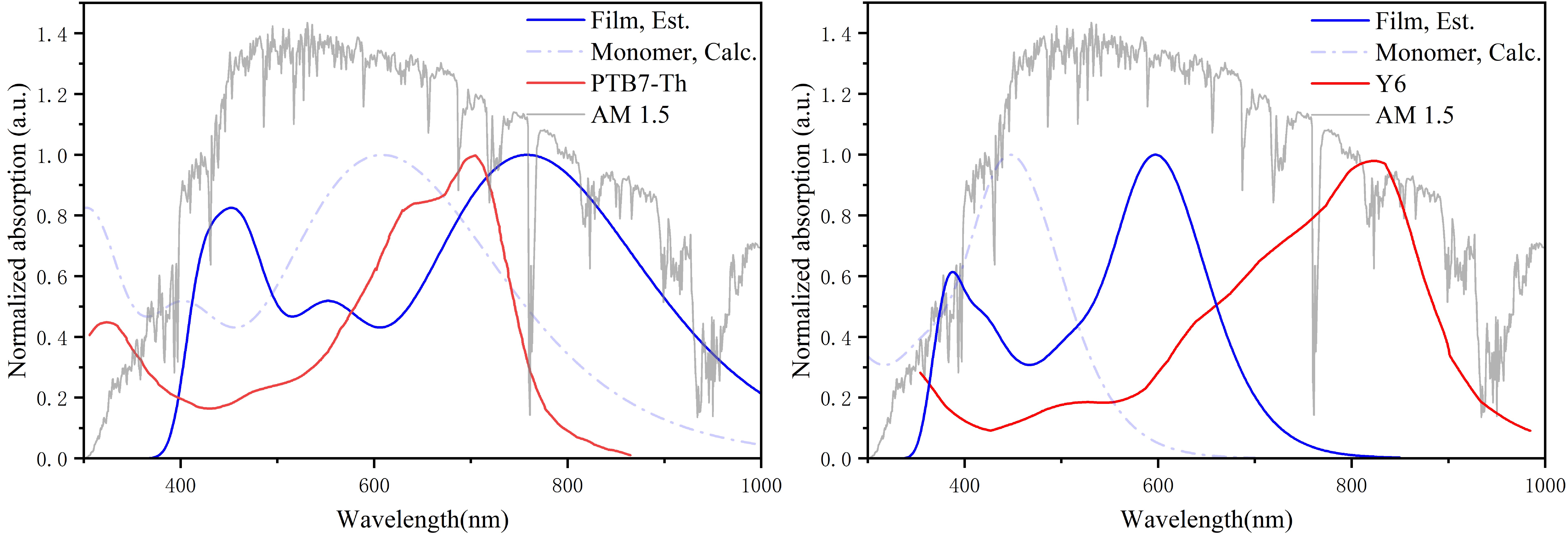}
    \caption{Spectral complementarity of generated OPV molecules. (a) Normalized absorption spectra for the donor PTB7-Th (red) and the corresponding generated acceptor. (b) Spectra for the acceptor Y6 (red) and the corresponding generated donor. In both panels, the calculated monomer absorption (Monomer, Calc.) of the generated molecule is shown alongside its estimated solid-state film spectrum (Film, Est.). The estimated film absorption successfully complements the absorption window of its partner, enabling broad coverage of the AM 1.5 solar spectrum (grey).}\label{fig:uv-vis}
\end{figure}

To validate the practical effectiveness of our framework, we selected representative candidates from two design tasks for quantum chemistry analysis. All DFT and TD-DFT calculations were performed using the Gaussian 16 and GaussView 6 software package \cite{g16,gv6}. In organic photovoltaics, two factors are critical for high performance: (1) Energy Level Alignment and (2) Spectral Complementarity. Proper alignment of the donor-acceptor frontier molecular orbitals (HOMO/LUMO) is essential to create a driving force for separating light-induced excitons into free charges. Concurrently, the donor and acceptor should absorb different regions of the solar spectrum, working in concert to achieve panchromatic light harvesting and maximize current generation.
Our framework demonstrated success in optimizing both factors, as illustrated in Figure~\ref{fig:uv-vis}. The figure displays the calculated monomer absorption spectra (Monomer, Calc.) for our representative generated molecules and their estimated solid-state film spectra (Film, Est.). Guided by experimental literature \cite{Y6,Y6-2,PTB7}, which typically reports a 100-200 nm redshift from solution to solid-state films due to increased molecular planarity and strong intermolecular $\pi - \pi$ interactions, we selected a representative shift of 150 nm for our estimation.  For the PTB7-Th pair, our generated acceptor extends the absorption deep into the near-infrared region (700-900 nm), perfectly complementing the donor's visible light absorption. Conversely, for the Y6 pair, our generated donor effectively fills the visible spectrum gap (400-650 nm) left by the NIR-absorbing Y6. In addition to this ideal spectral complementarity, quantum chemistry calculations confirmed that both pairs possess well-aligned frontier molecular orbital energies conducive to efficient exciton dissociation. This dual validation of optical and electronic properties underscores our framework's ability to discover novel, high-potential donor-acceptor partners.

\section{Discussion}
The GNN-based model provides interpretable predictions by linking molecular features to photovoltaic performance, with cross-attention highlighting donor–acceptor relationships aligned with known principles. However, real-world discrepancies arise due to fabrication variability, molecular stability, and synthetic feasibility, which are not fully captured by theoretical models. While adding fabrication data or quantum-level accuracy could improve predictions, it would significantly increase complexity. To maintain efficiency, our reinforcement learning framework relies on SMILES representations for practical donor–acceptor screening. Future directions include enhancing model generalization via active learning and domain adaptation, embedding experimental constraints directly into the model, and refining the reward function to better balance competing goals like performance, synthetic accessibility, and sustainability.

\section{Conclusion}
This study introduces a dual-pronged machine learning framework combining accurate performance prediction with efficient molecular generation to accelerate OPV material discovery. We constructed the large-scale OPV\textsuperscript{2}D dataset and employed the OPV Classifier (OPVC) for activity screening, alongside a multi-task GNN comprising MOE\textsuperscript{2} and P\textsuperscript{3} for precise prediction of frontier orbital energies and PCE. In parallel, MatGPT, fine-tuned via reinforcement learning, was used to generate candidate molecules with both high performance and validity. Experimental results demonstrate superior predictive accuracy and molecule quality, offering a powerful tool for rapid OPV screening. Future work will integrate experimental constraints and synthetic conditions to improve generalizability and real-world relevance for sustainable energy development.

\section{Acknowledgments}
This work is supported by the National Key R\&D Program of China (No. 2022ZD0117501), the Scientific Research Innovation Capability Support Project for Young Faculty (ZYGXQNJSKYCXNLZCXM-E7), the Tsinghua University Initiative Scientific Research Program (Student Academic Research Advancement Program: Zhuiguang Special Project) (Grant No. 20257020001), as well as the Tsinghua University Initiative Scientific Research Program and the Carbon Neutrality and Energy System Transformation (CNEST) Program led by Tsinghua University.

\bibliography{aaai2026}

@article{su2024roformer,
  title={Roformer: Enhanced transformer with rotary position embedding},
  author={Su, Jianlin and Ahmed, Murtadha and Lu, Yu and Pan, Shengfeng and Bo, Wen and Liu, Yunfeng},
  journal={Neurocomputing},
  volume={568},
  pages={127063},
  year={2024},
  publisher={Elsevier}
}

@article{brody2021attentive,
  title={How attentive are graph attention networks?},
  author={Brody, Shaked and Alon, Uri and Yahav, Eran},
  journal={arXiv preprint arXiv:2105.14491},
  year={2021}
}

@incollection{shockley2018detailed,
  title={Detailed balance limit of efficiency of p--n junction solar cells},
  author={Shockley, William and Queisser, Hans},
  booktitle={Renewable energy},
  pages={Vol2\_35--Vol2\_54},
  year={2018},
  publisher={Routledge}
}

@article{Sun2024,
author = {Sun, Jinyu and Li, Dongxu and Zou, Jie and Zhu, Shaofeng and Xu, Cong and Zou, Yingping and Zhang, Zhimin and Lu, Hongmei},
title = {Accelerating the discovery of acceptor materials for organic solar cells by deep learning},
journal = {npj Computational Materials},
volume = {10},
number = {1},
pages = {181},
year = {2024},
type = {Journal Article}
}

@article{Hutchison2023,
   author = {Greenstein, Brianna L. and Hutchison, Geoffrey R.},
   title = {Screening Efficient Tandem Organic Solar Cells with Machine Learning and Genetic Algorithms},
   journal = {The Journal of Physical Chemistry C},
   volume = {127},
   number = {13},
   pages = {6179-6191},
   year = {2023},
   type = {Journal Article}
}

@article{Lopez201751k,
   author = {Lopez, Steven A. and Sanchez-Lengeling, Benjamin and de Goes Soares, Julio and Aspuru-Guzik, Alán},
   title = {Design Principles and Top Non-Fullerene Acceptor Candidates for Organic Photovoltaics},
   journal = {Joule},
   volume = {1},
   number = {4},
   pages = {857-870},
   year = {2017},
   type = {Journal Article}
}

@article{Saeki2021,
   author = {Miyake, Yuta and Saeki, Akinori},
   title = {Machine Learning-Assisted Development of Organic Solar Cell Materials: Issues, Analyses, and Outlooks},
   journal = {The Journal of Physical Chemistry Letters},
   volume = {12},
   number = {51},
   pages = {12391-12401},
   year = {2021},
   type = {Journal Article}
}

@article{Beyond2024review,
   author = {Seifrid, Martin and Lo, Stanley and Choi, Dylan G. and Tom, Gary and Le, My Linh and Li, Kunyu and Sankar, Rahul and Vuong, Hoai-Thanh and Wakidi, Hiba and Yi, Ahra and Zhu, Ziyue and Schopp, Nora and Peng, Aaron and Luginbuhl, Benjamin R. and Nguyen, Thuc-Quyen and Aspuru-Guzik, Alán},
   title = {Beyond molecular structure: critically assessing machine learning for designing organic photovoltaic materials and devices},
   journal = {Journal of Materials Chemistry A},
   volume = {12},
   number = {24},
   pages = {14540-14558},
   year = {2024},
   type = {Journal Article}
}

@article{Min2020,
   author = {Wu, Yao and Guo, Jie and Sun, Rui and Min, Jie},
   title = {Machine learning for accelerating the discovery of high-performance donor/acceptor pairs in non-fullerene organic solar cells},
   journal = {npj Computational Materials},
   volume = {6},
   number = {1},
   pages = {120},
   year = {2020},
   type = {Journal Article}
}

@article{ZhangSA2025,
author = {Zhang, Shizhao and Li, Shuixing and Song, Siqin and Zhao, Yang and Gao, Liang and Chen, Hongzheng and Li, Hanying and Lin, Jiaping},
title = {Deep Learning-Assisted Design of Novel Donor–Acceptor Combinations for Organic Photovoltaic Materials with Enhanced Efficiency},
journal = {Advanced Materials},
volume = {37},
number = {4},
pages = {2407613},
year = {2025}
}

@article{Y6,
title = {Single-Junction Organic Solar Cell with over 15\% Efficiency Using Fused-Ring Acceptor with Electron-Deficient Core},
journal = {Joule},
volume = {3},
number = {4},
pages = {1140-1151},
year = {2019},
author = {Jun Yuan and Yunqiang Zhang and Liuyang Zhou and Guichuan Zhang and Hin-Lap Yip and Tsz-Ki Lau and Xinhui Lu and Can Zhu and Hongjian Peng and Paul A. Johnson and Mario Leclerc and Yong Cao and Jacek Ulanski and Yongfang Li and Yingping Zou},

}

@article{1060,
   author = {Zhang, Wenlin and Zou, Yurong and Wang, Xin and Chen, Junxian and Xu, Dingguo},
   title = {Deep learning accelerated high-throughput screening of organic solar cells},
   journal = {Journal of Materials Chemistry C},
   volume = {13},
   number = {10},
   pages = {5295-5306},
   ISSN = {2050-7526},
   DOI = {10.1039/D5TC00111K},
   url = {http://dx.doi.org/10.1039/D5TC00111K},
   year = {2025},
   type = {Journal Article}
}

@article{300,
   author = {Das, Bibhas and Mondal, Anirban},
   title = {Predictive Modeling and Design of Organic Solar Cells: A Data-Driven Approach for Material Innovation},
   journal = {ACS Applied Energy Materials},
   volume = {7},
   number = {20},
   pages = {9349-9363},
   note = {doi: 10.1021/acsaem.4c01847},
   DOI = {10.1021/acsaem.4c01847},
   url = {https://doi.org/10.1021/acsaem.4c01847},
   year = {2024},
   type = {Journal Article}
}

@article{100,
   author = {Khatua, Rudranarayan and Das, Bibhas and Mondal, Anirban},
   title = {Physics-Informed Machine Learning with Data-Driven Equations for Predicting Organic Solar Cell Performance},
   journal = {ACS Applied Materials \& Interfaces},
   volume = {16},
   number = {42},
   pages = {57467-57480},
   note = {doi: 10.1021/acsami.4c10868},
   ISSN = {1944-8244},
   DOI = {10.1021/acsami.4c10868},
   url = {https://doi.org/10.1021/acsami.4c10868},
   year = {2024},
   type = {Journal Article}
}

@article{sterling2015zinc,
  title={ZINC 15--ligand discovery for everyone},
  author={Sterling, Teague and Irwin, John J},
  journal={Journal of chemical information and modeling},
  volume={55},
  number={11},
  pages={2324--2337},
  year={2015},
  publisher={ACS Publications}
}

@article{gaulton2012chembl,
  title={ChEMBL: a large-scale bioactivity database for drug discovery},
  author={Gaulton, Anna and Bellis, Louisa J and Bento, A Patricia and Chambers, Jon and Davies, Mark and Hersey, Anne and Light, Yvonne and McGlinchey, Shaun and Michalovich, David and Al-Lazikani, Bissan and others},
  journal={Nucleic acids research},
  volume={40},
  number={D1},
  pages={D1100--D1107},
  year={2012},
  publisher={Oxford University Press}
}

@article{radford2019language,
  title={Language models are unsupervised multitask learners},
  author={Radford, Alec and Wu, Jeffrey and Child, Rewon and Luan, David and Amodei, Dario and Sutskever, Ilya and others},
  journal={OpenAI blog},
  volume={1},
  number={8},
  pages={9},
  year={2019}
}

@article{polykovskiy2020molecular,
  title={Molecular sets (MOSES): a benchmarking platform for molecular generation models},
  author={Polykovskiy, Daniil and Zhebrak, Alexander and Sanchez-Lengeling, Benjamin and Golovanov, Sergey and Tatanov, Oktai and Belyaev, Stanislav and Kurbanov, Rauf and Artamonov, Aleksey and Aladinskiy, Vladimir and Veselov, Mark and others},
  journal={Frontiers in pharmacology},
  volume={11},
  pages={565644},
  year={2020},
  publisher={Frontiers Media SA}
}

@article{brown1992class,
  title={Class-based n-gram models of natural language},
  author={Brown, Peter F and Della Pietra, Vincent J and Desouza, Peter V and Lai, Jennifer C and Mercer, Robert L},
  journal={Computational linguistics},
  volume={18},
  number={4},
  pages={467--480},
  year={1992}
}

@article{gomez2018automatic,
  title={Automatic chemical design using a data-driven continuous representation of molecules},
  author={G{\'o}mez-Bombarelli, Rafael and Wei, Jennifer N and Duvenaud, David and Hern{\'a}ndez-Lobato, Jos{\'e} Miguel and S{\'a}nchez-Lengeling, Benjam{\'\i}n and Sheberla, Dennis and Aguilera-Iparraguirre, Jorge and Hirzel, Timothy D and Adams, Ryan P and Aspuru-Guzik, Al{\'a}n},
  journal={ACS central science},
  volume={4},
  number={2},
  pages={268--276},
  year={2018},
  publisher={ACS Publications}
}

@MISC{rdkit,
  title = {{RDK}it: Open-source cheminformatics},
  howpublished = {\url{http://www.rdkit.org}},
  note = {[Online; accessed 11-April-2013]},
  key = {RDKit, online},
  year = {2013}
}

@article{kim2025pubchem,
  title={PubChem 2025 update},
  author={Kim, Sunghwan and Chen, Jie and Cheng, Tiejun and Gindulyte, Asta and He, Jia and He, Siqian and Li, Qingliang and Shoemaker, Benjamin A and Thiessen, Paul A and Yu, Bo and others},
  journal={Nucleic acids research},
  volume={53},
  number={D1},
  pages={D1516--D1525},
  year={2025},
  publisher={Oxford University Press}
}

@article{hendrycks2016gaussian,
  title={Gaussian error linear units (gelus)},
  author={Hendrycks, Dan and Gimpel, Kevin},
  journal={arXiv preprint arXiv:1606.08415},
  year={2016}
}

@article{holtzman2019curious,
  title={The curious case of neural text degeneration},
  author={Holtzman, Ari and Buys, Jan and Du, Li and Forbes, Maxwell and Choi, Yejin},
  journal={arXiv preprint arXiv:1904.09751},
  year={2019}
}

@article{breiman2001random,
  title={Random forests},
  author={Breiman, Leo},
  journal={Machine learning},
  volume={45},
  number={1},
  pages={5--32},
  year={2001},
  publisher={Springer}
}

@article{lin1992self,
  title={Self-improving reactive agents based on reinforcement learning, planning and teaching},
  author={Lin, Long-Ji},
  journal={Machine learning},
  volume={8},
  number={3},
  pages={293--321},
  year={1992},
  publisher={Springer}
}

@article{kipf2016semi,
  title={Semi-Supervised Classification with Graph Convolutional Networks},
  author={Kipf, TN},
  journal={arXiv preprint arXiv:1609.02907},
  year={2016}
}

@article{xu2018powerful,
  title={How powerful are graph neural networks?},
  author={Xu, Keyulu and Hu, Weihua and Leskovec, Jure and Jegelka, Stefanie},
  journal={arXiv preprint arXiv:1810.00826},
  year={2018}
}

@article{cortes1995support,
  title={Support-vector networks},
  author={Cortes, Corinna and Vapnik, Vladimir},
  journal={Machine learning},
  volume={20},
  number={3},
  pages={273--297},
  year={1995},
  publisher={Springer}
}

@article{friedman2001greedy,
  title={Greedy function approximation: a gradient boosting machine},
  author={Friedman, Jerome H},
  journal={Annals of statistics},
  pages={1189--1232},
  year={2001},
  publisher={JSTOR}
}

@inproceedings{chen2016xgboost,
  title={Xgboost: A scalable tree boosting system},
  author={Chen, Tianqi and Guestrin, Carlos},
  booktitle={Proceedings of the 22nd acm sigkdd international conference on knowledge discovery and data mining},
  pages={785--794},
  year={2016}
}

@article{prokhorenkova2018catboost,
  title={CatBoost: unbiased boosting with categorical features},
  author={Prokhorenkova, Liudmila and Gusev, Gleb and Vorobev, Aleksandr and Dorogush, Anna Veronika and Gulin, Andrey},
  journal={Advances in neural information processing systems},
  volume={31},
  year={2018}
}

@article{ke2017lightgbm,
  title={Lightgbm: A highly efficient gradient boosting decision tree},
  author={Ke, Guolin and Meng, Qi and Finley, Thomas and Wang, Taifeng and Chen, Wei and Ma, Weidong and Ye, Qiwei and Liu, Tie-Yan},
  journal={Advances in neural information processing systems},
  volume={30},
  year={2017}
}

@article{seeger2004gaussian,
  title={Gaussian processes for machine learning},
  author={Seeger, Matthias},
  journal={International journal of neural systems},
  volume={14},
  number={02},
  pages={69--106},
  year={2004},
  publisher={World Scientific}
}

@article{pedregosa2011scikit,
  title={Scikit-learn: Machine learning in Python},
  author={Pedregosa, Fabian and Varoquaux, Ga{\"e}l and Gramfort, Alexandre and Michel, Vincent and Thirion, Bertrand and Grisel, Olivier and Blondel, Mathieu and Prettenhofer, Peter and Weiss, Ron and Dubourg, Vincent and others},
  journal={the Journal of machine Learning research},
  volume={12},
  pages={2825--2830},
  year={2011},
  publisher={JMLR. org}
}

@inproceedings{duan2020ngboost,
  title={Ngboost: Natural gradient boosting for probabilistic prediction},
  author={Duan, Tony and Anand, Avati and Ding, Daisy Yi and Thai, Khanh K and Basu, Sanjay and Ng, Andrew and Schuler, Alejandro},
  booktitle={International conference on machine learning},
  pages={2690--2700},
  year={2020},
  organization={PMLR}
}

@article{guimaraes2017objective,
  title={Objective-reinforced generative adversarial networks (organ) for sequence generation models},
  author={Guimaraes, Gabriel Lima and Sanchez-Lengeling, Benjamin and Outeiral, Carlos and Farias, Pedro Luis Cunha and Aspuru-Guzik, Al{\'a}n},
  journal={arXiv preprint arXiv:1705.10843},
  year={2017}
}

@article{radford2018improving,
  title={Improving language understanding by generative pre-training},
  author={Radford, Alec and Narasimhan, Karthik and Salimans, Tim and Sutskever, Ilya and others},
  year={2018},
  publisher={San Francisco, CA, USA}
}

@article{weininger1988smiles,
  title={SMILES, a chemical language and information system. 1. Introduction to methodology and encoding rules},
  author={Weininger, David},
  journal={Journal of chemical information and computer sciences},
  volume={28},
  number={1},
  pages={31--36},
  year={1988},
  publisher={ACS Publications}
}

@article{Y6-2,
author = {P Christopholi, Leticia and Marchiori, Cleber F. N. and Jalan, Ishita and Opitz, Andreas and Muntean, Stela Andrea and Moons, Ellen},
title = {Role of the Solvent on the Orientation of Y-Type Acceptor Molecules in Spin-Coated Films},
journal = {The Journal of Physical Chemistry C},
volume = {128},
number = {42},
pages = {17825-17835},
year = {2024},
doi = {10.1021/acs.jpcc.4c04176},
}

@article{PTB7,
title = {Small bandgap non-fullerene acceptor enables efficient PTB7-Th solar cell with near 0 eV HOMO offset},
journal = {Journal of Energy Chemistry},
volume = {52},
pages = {60-66},
year = {2021},
issn = {2095-4956},
doi = {https://doi.org/10.1016/j.jechem.2020.03.058},
url = {https://www.sciencedirect.com/science/article/pii/S209549562030190X},
author = {Chao Li and Qihui Yue and Hao Wu and Baolin Li and Haijun Fan and Xiaozhang Zhu},
}

@misc{g16,
author={M. J. Frisch and G. W. Trucks and H. B. Schlegel and G. E. Scuseria and M. A. Robb and J. R. Cheeseman and G. Scalmani and V. Barone and G. A. Petersson and H. Nakatsuji and X. Li and M. Caricato and A. V. Marenich and J. Bloino and B. G. Janesko and R. Gomperts and B. Mennucci and H. P. Hratchian and J. V. Ortiz and A. F. Izmaylov and J. L. Sonnenberg and D. Williams-Young and F. Ding and F. Lipparini and F. Egidi and J. Goings and B. Peng and A. Petrone and T. Henderson and D. Ranasinghe and V. G. Zakrzewski and J. Gao and N. Rega and G. Zheng and W. Liang and M. Hada and M. Ehara and K. Toyota and R. Fukuda and J. Hasegawa and M. Ishida and T. Nakajima and Y. Honda and O. Kitao and H. Nakai and T. Vreven and K. Throssell and Montgomery, {Jr.}, J. A. and J. E. Peralta and F. Ogliaro and M. J. Bearpark and J. J. Heyd and E. N. Brothers and K. N. Kudin and V. N. Staroverov and T. A. Keith and R. Kobayashi and J. Normand and K. Raghavachari and A. P. Rendell and J. C. Burant and S. S. Iyengar and J. Tomasi and M. Cossi and J. M. Millam and M. Klene and C. Adamo and R. Cammi and J. W. Ochterski and R. L. Martin and K. Morokuma and O. Farkas and J. B. Foresman and D. J. Fox},
title={Gaussian˜16 {R}evision {C}.01},
year={2016},
note={Gaussian Inc. Wallingford CT}
}

@misc{gv6,
author={Roy Dennington and Todd A. Keith and John M. Millam},
title={GaussView {V}ersion {6}},
note={Semichem Inc. Shawnee Mission KS},
year={2019}
}

\end{document}